\documentclass[reprint,amsmath,amssymb,aps,prl,superscriptaddress, showpacs, eprint, longbibliography]{revtex4-1}

\usepackage{graphicx}
\usepackage{bm}
\usepackage{siunitx}
\usepackage{physics}
\usepackage{hyperref}

\renewcommand{\vec}{\bm}
\newcommand{\kB}{k_{\mathrm{B}}}
\newcommand{\ani}[1]{\hat{#1}}
\newcommand{\cre}[1]{\hat{#1}^\dagger}
\newcommand{\ene}{\varepsilon}

\newcommand{\Lg}{\mathcal{L}}
\newcommand{\nablak}{\nabla_{\vec{k}}}
\newcommand{\omegap}{\omega_{+}}
\newcommand{\chie}{\chi_{e}}
\newcommand{\chit}{\chi_{t}}

\newcommand{\tauord}{T_{\tau}}
\newcommand{\G}{\mathcal{G}}
\newcommand{\GR}{G^R}
\newcommand{\GA}{G^A}

\begin{document}

\title{
Theory of thermopolarization effect
}

\author{Yugo Onishi}
\affiliation{Department of Physics, Massachusetts Institute of Technology, Cambridge, MA 02139, USA}
% \email{yugo0o24@mit.edu}

\author{Hiroki Isobe}
\affiliation{
	Center for Emergent Matter Science (CEMS), RIKEN, Wako, Saitama 351-0198, Japan
}
\affiliation{
	Department of Physics, Kyushu University, Fukuoka 819-0395, Japan
}

\author{Atsuo Shitade}
\affiliation{
Institute of Scientific and Industrial Research, Osaka University, 8-1 Mihogaoka, Ibaraki, Osaka 567-0047, Japan
}

\author{Naoto Nagaosa}
\affiliation{
Center for Emergent Matter Science (CEMS), RIKEN, Wako, Saitama 351-0198, Japan
}
\affiliation{
	Fundamental Quantum Science Program, TRIP Headquarters, RIKEN, Wako 351-0198, Japan
}

\date{\today}
\begin{abstract}
We study the polarization response to the temperature gradient in insulators, known as the thermopolarization effect. We show that this response can be understood through the free energy response function to an electric field gradient, which we call $Q$-tensor. By using the $Q$-tensor, we present a unified description of the polarization responses to both electric fields and temperature gradients and derive the generalized Mott relation. Additionally, we draw an analogy with the anomalous Hall and Nernst effects. These effects are observable as the Seebeck effect where the linear size of the system is shorter than the screening length.
\end{abstract}

\maketitle

Responses of insulators can reveal interesting nature of their ground states.
Despite the absence of itinerant electrons, insulators still can exhibit electric responses which reflect the quantum mechanical nature of the electronic states. A prominent example is the quantum Hall effect~\cite{Thouless1982,niu1985,vonklitzing1986, stormer1999,chang2013, park2023, park2023,chang2023} where DC Hall conductivity remains finite in insulators and is topologically quantized due to the Berry phase in $k$-space.
Interestingly, Berry phase effects can be probed also by statistical forces so that the Hall responses satisfy the Einstein and Mott relations~\cite{Xiao2006, Qin2011}. Additionally, the connection of Hall conductivity to magnetization~\cite{streda1983} allows a thermodynamic understanding of the quantum Hall effect.

Another important quantum feature beside the phase is superposition of states. For example, polarization response in insulators is described quantum mechanically~\cite{Resta1992,Resta1994} by the hybridization of the ground and excited states, namely, superposition. In contrast to Berry phase effects which often require the inversion and/or time reversal symmetry breaking, superposition phenomena occur under any symmetry, thus are more ubiquitous. Recent studies also relate the polarization response to quantum geometry~\cite{marzari1997,souza2000,resta2006,onishi2024_qw_arxiv,komissarov2024}, complementing the topological aspects related to Hall conductance in insulators.

A natural question is then whether polarization involving wavefunction hybridization can be induced by statistical forces such as a temperature gradient. The thermopolarization effect, where a temperature gradient induces polarization, was previously linked to ionic displacement and charge redistribution~\cite{Marvan1969, Tagantsev1987,Tagantsev1991}. However, it remains unclear whether quantum mechanical electrons can develop superposition and hence polarization through statistical forces, as well as their relation to thermodynamics.

In this work, we study polarization responses in quantum electronic insulators, especially the thermopolarization effect, described by:
\begin{align}
	\vec{P} &= \chie \vec{E} + \chit\qty(-\frac{\nabla T}{T}), \label{eq:polarization_response}
\end{align}
where $\vec{P}$ is the induced polarization density, $\vec{E}$ is the electric field, and $T$ is the temperature. $\chie$ is the electric susceptibility (or also called polarizability), and $\chit$ is the thermopolarizability characterizing the thermopolarization effect, the main topic of this work. We show that a temperature gradient induces finite polarization in insulators, albeit exponentially small at low temperatures. The thermopolarization effect results in a finite voltage drop across the system under a certain condition (clarified below), and hence it contributes to the Seebeck effect. 
These are shown through a unified thermodynamic description of the polarization response~\eqref{eq:polarization_response}
analogous to magnetization in the quantum Hall effect, which further leads to the generalized Mott relation between electric susceptibility and thermopolarizability. Our results highlight a new aspect of quantum electronic insulators and its relation to thermodynamics.

% Furthermore, we show that the polarization response~\eqref{eq:polarization_response} can be understood thermodynamically in a unified manner 

We begin by defining polarization at finite temperatures considered in this work. We consider polarization in local equilibrium states under external stimuli, such as an electric field or a temperature gradient. In local equilibrium, the system is described by the local free energy density $F(\vec{r})$, and the polarization density $\vec{P}(\vec{r})$ is defined as the response of the free energy density to the electric field:
\begin{align}
	\Delta{F}(\vec{r}) = -P^\alpha(\vec{r}){E_{\alpha}(\vec{r})}. \label{eq:def_polarization}
\end{align}
While spontaneous polarization in equilibrium can be subtle in periodic systems~\cite{Resta1992, Resta1994}, our focus in this work is rather on the \textit{induced} polarization by the external stimuli. The induced polarization is well-defined and observable in periodic systems. We note that the polarization studied in this work is purely electronic and we will neglect phonon effects, although thermopolarization in real materials may involve thermally activated phonons through electron-phonon coupling.

The polarization induced by an electric field or a temperature gradient is characeterized by the electric susceptibility $\chie$ and thermopolarizability $\chit$ in Eq.~\eqref{eq:polarization_response}. 
% The electric susceptibility $\chie$ characterizes the polarization response to the electric field, while the thermopolarizability $\chit$ characterizes the polarization response to the temperature gradient. 
By taking the derivative of Eq.~\eqref{eq:def_polarization}, we see that $\chie, \chit$ are given by the second derivative of the free energy density with respect to the electric field and the temperature gradient, respectively:
\begin{align}
	\chie^{\alpha\beta} &= -\pdv[2]{F}{E_{\alpha}}{E_{\beta}}, \label{eq:polarization_response_electric} \\
	\chit^{\alpha\beta} &= T\pdv[2]{F}{E_{\alpha}}{(\partial_{\beta}T)}. \label{eq:polarization_response_temperature}
\end{align}
Here, $\partial_{\alpha}$ denotes the derivative with respect to the spatial coordinate $r_{\alpha}$.

To understand the polarization response $\chie$ and $\chit$, it is convenient to consider the response of the free energy to the gradient of the electric field, which we denote by $Q^{\alpha\beta}$~\cite{Daido2020}: 
\begin{align}
	\pdv{F}{(\partial_\beta E_{\alpha})} = - Q^{\alpha\beta}. \label{eq:Q-tensor_def}
\end{align}
While $Q^{\alpha\beta}$ is called a thermodynamic quadrupole moment in Ref.~\cite{Daido2020}, to avoid confusion with the other definitions of quadrupole moment in literature~\cite{ren2021,ono2019,wheeler2019,kang2019b}, we will refer to $Q^{\alpha\beta}$ as $Q$-tensor in this work. 

The $Q$-tensor defined by Eq.~\eqref{eq:Q-tensor_def} is closely related to polarization response. To see this, consider a system with chemical potential $\mu_0$ and temperature $T_0$, and apply a scalar potential $\Delta\phi(\vec{r})$ and a spatially varying temperature $T(\vec{r})=T_0 + \Delta T(\vec{r})$ to the system. Since they result in an electric field $E_{\alpha}=-\partial_{\alpha}(\Delta\phi)$ and a temperature gradient $\partial_{\alpha} T=\partial_{\alpha}(\Delta T)$, the change of the free energy density due to the electric field and the temperature gradient is then given by 
using Eqs.~\eqref{eq:polarization_response_electric}, \eqref{eq:polarization_response_temperature} as
\begin{align}
	\Delta {F(\vec{r})} = -\frac{\chie^{\alpha\beta}}{2} E_{\alpha} E_{\beta} + \frac{\chit^{\alpha\beta}}{T} E_{\alpha} (\partial_{\beta}T), \label{eq:free_energy_change}
\end{align}
where we omitted terms of the first order in $E_{\alpha}$ and $\partial_{\beta}T$, such as terms involving the spontaneous polarization, because they are irrelevant to the \textit{induced} polarization~\eqref{eq:polarization_response}.
By partially integrating Eq.~\eqref{eq:free_energy_change}, we obtain~\footnote{We can neglect the surface contribution in partial integral because its contribution is much smaller than $\Delta{F}(\vec{r})$.} 
\begin{align}
	\Delta{F}(\vec{r}) = -\qty(\frac{\chie^{\alpha\beta}}{2} \Delta \phi(\vec{r}) + \frac{\chit^{\alpha\beta}}{T} \Delta T(\vec{r})) (\partial_{\beta}E_{\alpha}). \label{eq:free_energy_change2}
\end{align}
Comparing Eq.~\eqref{eq:free_energy_change2} with Eq.~\eqref{eq:Q-tensor_def}, we find the change in the $Q$-tensor $\Delta{Q}$ due to the potential and the temperature variation as
\begin{align}
	\Delta{Q^{\alpha\beta}(\vec{r})} &= \chie^{\alpha\beta}\Delta\phi(\vec{r}) + \frac{\chit^{\alpha\beta}}{T} \Delta{T(\vec{r})}. \label{eq:Q-tensor_chie_chiT}
\end{align}
Note that the factor of $1/2$ in Eq.~\eqref{eq:free_energy_change2} is canceled in Eq.~\eqref{eq:Q-tensor_chie_chiT} as $\Delta \phi$ and $\partial_{\beta}E_{\alpha}=-\partial_{\beta}\partial_{\alpha}(\Delta\phi)$ are not independent.

The relation~\eqref{eq:Q-tensor_chie_chiT} can be viewed as a relation between the change of $Q$-tensor and that of polarization: 
\begin{align}
	\Delta{P^\alpha(\vec{r})} = -\partial_{\beta}(\Delta Q^{\alpha\beta}(\vec{r})). \label{eq:P-Q}
\end{align}
This is obtained by taking the derivative of Eq.~\eqref{eq:Q-tensor_def} with respect to the position $r_{\beta}$ and using Eq.~\eqref{eq:polarization_response}. To emphasize the polarization appearing here is the \textit{induced} polarization, we denote it by $\Delta{P}$, although it is simply denoted as $\vec{P}$ in Eq.~\eqref{eq:polarization_response}.

Eq.~\eqref{eq:P-Q} shows that the spatial change of the $Q$-tensor results in the induced polarization. 
This is analogous to the relation between charge density $\rho$ and polarization $\vec{P}$, or current $\vec{j}$ and magnetization $\vec{M}$: $\rho=-\div \vec{P}$, $\vec{j} = \curl\vec{M}$. Intuitively, the $Q$-tensor can be viewed as a ``pair'' of opposite polarizations in the system, as polarization represents a ``pair'' of opposite charges. When the $Q$-tensor spatially varies, the polarization is no longer fully cancelled and the gradient of the $Q$-tensor appears as the polarization. The $Q$-tensor can vary spatially either by an external potential $\phi$ or a temperature gradient $T$ as in Eq.~\eqref{eq:Q-tensor_chie_chiT}. 
It is important to note that the relation~\eqref{eq:P-Q} applies only to the \textit{induced} polarization $\Delta{\vec{P}}$. In contrast, spontaneous polarization can depend on the boundary conditions and is not necessarily related to the $Q$-tensor, which is a bulk quantity.

The quantum mechanical expression for the $Q$-tensor in noninteracting electronic systems was previously obtained~\cite{Daido2020}, from which we find the $Q$-tensor of insulators as
\begin{align}
	Q^{\alpha\beta}(\mu, T) &= e\sum_n \int[\dd{k}] g_n^{\alpha\beta} f(\ene_n-\mu) - \frac{1}{e} \int_{-\infty}^\mu\dd{\mu'} \chi_0^{\alpha\beta}(\mu', T) \label{eq:quadrupole}
\end{align}
where $\mu$ is the chemical potential, $T$ is the temperature, $[\dd{k}]=\dd^d k/(2\pi)^d$ with the spatial dimension $d$, $\ene_n = \ene_n(\vec{k})$ is the $n$-th band dispersion, $e(<0)$ is the electron charge, and $f(\ene)=(e^{\beta\ene}+1)^{-1}$ is the Fermi distribution function with $\beta=(\kB T)^{-1}$. $g_n^{\alpha\beta}=\sum_{m(\neq n)} \Re[A^\alpha_{nm}A^\beta_{mn}]$ is the quantum metric for $n$-th band, and $A^\alpha_{nm}=\mel{u_n}{i\pdv*{k_\alpha}}{u_m}$ is the interband Berry connection. $\chi_0^{\alpha\beta}(\mu, T)$ is the electric susceptibility at low temperatures:
\begin{align}
	\chi_0^{\alpha\beta}(\mu, T) &= e^2\int[\dd{k}]\sum_{n}G_n^{\alpha\beta} f(\ene_n-\mu). \label{eq:polarizability}
\end{align}
where $G_n^{\alpha\beta}=-2\sum_{m(\neq n)} \Re[A^\alpha_{nm}A^\beta_{mn}]/(\ene_n-\ene_m)$. At $T=0$, $\chi_0^{\alpha\beta}$ reduces to the well-known expression for electric susceptibility~\cite{Vanderbilt2018}. 

\begin{figure}[htbp]
	\centering
	\includegraphics[width=0.8\columnwidth]{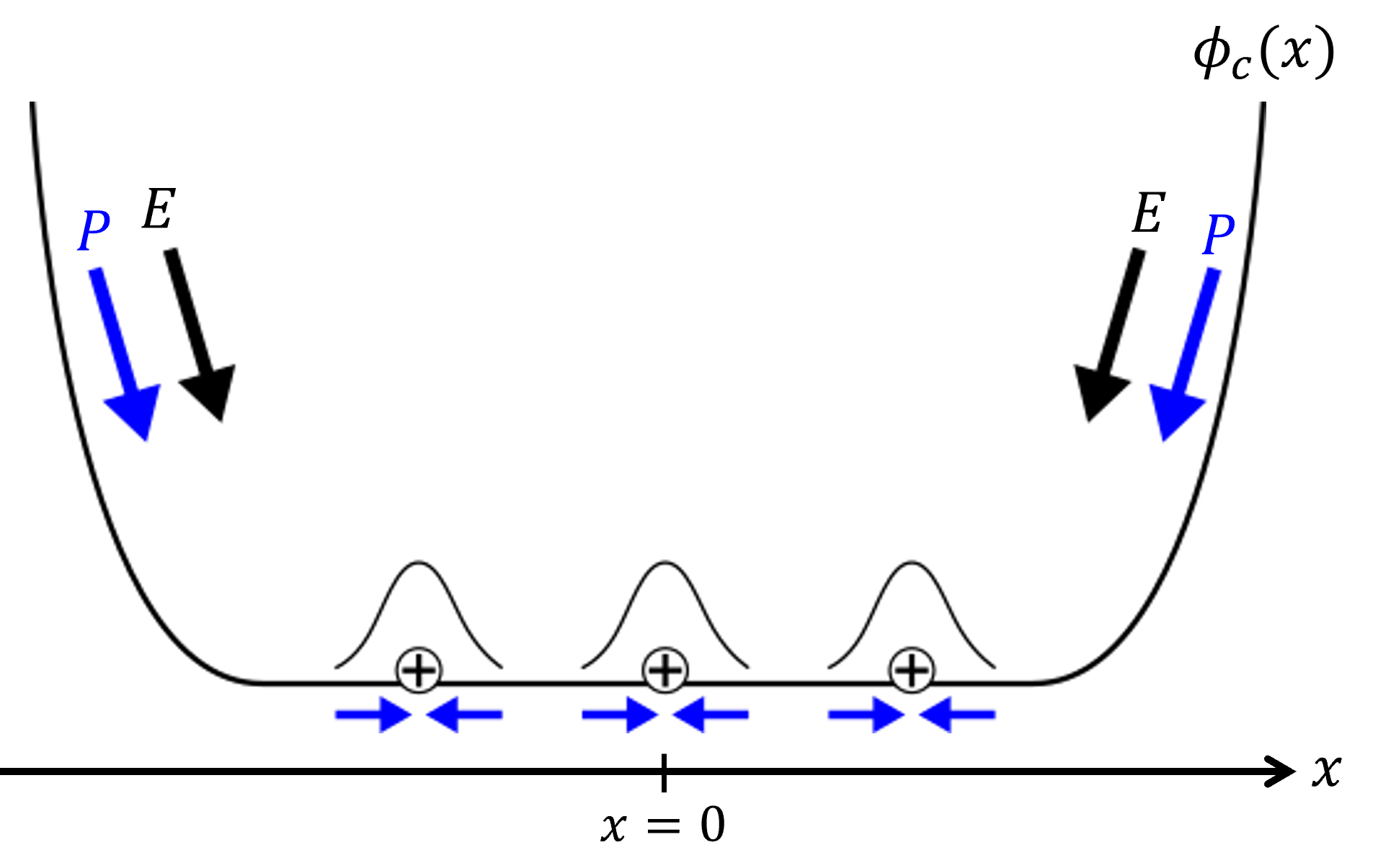}
	\caption{$Q$-tensor in insulators. The $Q$-tensor is given by the sum of the wave packet contribution and the edge contribution. The wave packet contribution comes from the spread of the wave packet around a positive ion, while the edge contribution is the polarization induced at the edge of the system due to the confining potential $\phi_c(x)$. }
	\label{fig:quadrupole}
\end{figure}

Each term in Eq.~\eqref{eq:quadrupole} can be interpreted analogously to orbital magnetization~\cite{Xiao2010} as illustrated schematically in Fig.~\ref{fig:quadrupole}. The first term in Eq.~\eqref{eq:quadrupole} is a ``wave packet'' contribution. The quantum metric quantifies the spatial spread of the corresponding Wannier function~\cite{marzari1997}, and this spread creates a ``pair'' of opposite polarizations around the positive ion charge in each unit cell. 
% thus naively the wave function in each unit cell has spread of $g^{\alpha\beta}_n$ around the positive ion charge, forming a ``pair'' of polarization with the opposite sign. 
One may also interpret the first term as the polarization fluctuation~\cite{souza2000,resta2006}. The integral of quantum metric at zero temperature is recently termed as quantum weight~\cite{onishi2024_qw_arxiv}, and is studied in the context of quantum geometry as well~\cite{tam2024,komissarov2024}.

The second term in Eq.~\eqref{eq:quadrupole} represents an ``edge'' contribution to the $Q$-tensor. For simplicity, consider a finite one-dimensional system with edges modeled by a confining potential $\phi_c(x)$, as shown in Fig.~\ref{fig:quadrupole}. For the right edge ($x>0$), $\phi_c(x)$ is zero deep inside the bulk (set to be $x=0$) and increases towards outside the sample ($x\to\infty$) as $\phi_c\to \infty$ so that there is no electron outside the sample. The gradient of $\phi_c(x)$ induces the polarization at the edge, contributing to the $Q$-tensor. When $\phi_c$ varies slowly, we can treat the system as local equilibrium and the total induced polarization at the right edge is given by 
\begin{align}
	\int_{\rm edge} \dd{x} P(x) &= \int_0^\infty \dd{x} (-\partial_x\phi_c(x)) \chi_0(\mu-e\phi_c(x)) \nonumber \\
	&= -\int_{-\infty}^\mu \frac{\dd{\mu'}}{e} \chi_0(\mu'). %
\end{align} 
At the left edge, the induced polarization has the same magnitude but opposite sign, and contributes to the $Q$-tensor together with the polarization at the right edge. This edge picture also explains why the $Q$-tensor explicitly depends on the chemical potential, even when the chemical potential is within the gap. A change of chemical potential within the gap does not affect the state of the bulk, but does affect the edge polarization. This is analogous to the edge current in quantum Hall insulators, where changing the chemical potential modifies the edge current and hence the magnetization~\cite{halperin1982, Xiao2010}. However, we emphasize that edge polarization does {\it not} require topological edge states, in contrast to the quantum Hall insulators. 

Although it can be understood through the edge contribution, $Q$-tensor is a bulk property, similar to magnetization in quantum Hall insulators. The edge polarization is a consequence of the finite $Q$-tensor of the \textit{bulk} states when a gradient of $\phi_c$ is present at the edges.
While the confining potential $\phi_c$ is assumed to vary slowly here, the edge polarization and its relation to $Q$-tensor is expected to be independent of details of edges because they are determined by bulk properties. 

We also note that the ``edge'' contribution in the $Q$-tensor describes an interband effect involving hybridization between occupied and unoccupied bands. While the ``wave packet'' contribution depends solely on the Bloch wavefunction of the occupied bands as the quantum metric is given by $g_{n}^{\alpha\beta}=\ip{\partial_{k_{\alpha}}u_n}{\partial_{k_{\beta}}u_n} - \ip{\partial_{k_{\alpha}}u_n}{u_n}\ip{u_n}{\partial_{k_{\beta}}u_n}$, the edge contribution cannot be reduced to a property of the occupied bands alone. 

As a simple example, consider strongly localized electrons in a periodic potential in one dimension. When electrons are strongly localized around potential minima, the system is well approximated as an array of harmonic oscillators. Neglecting electron hopping between unit cells, the band dispersion is flat and given by a single-particle energy of the harmonic oscillator, $\ene_\nu=\hbar\omega_0(\nu+1/2)$ with $\omega_0$ the frequency of the harmonic oscillators. When each unit cell contains one electron and only the lowest band ($\nu=0$) is fully filled, the electric susceptibility at $T=0$ is given by 
$\chi_{{\rm HO}, 0}=ne^2/(m\omega_0^2)$ with the electron density $n=1/a$,  the mass of an electron $m$, and the lattice constant $a$, while the quantum metric for $\nu=0$ is given by $g_0=x_0^2$ with $x_0^2=\hbar/(2m\omega_0)$ the zero point fluctuation of the oscillators. 
Therefore, the $Q$-tensor at $T=0$ is then given by 
\begin{align}
	Q_{\rm HO} = enx_0^2 - \frac{\mu-\ene_0}{e}\chi_{{\rm HO}, 0}.
\end{align}
As described above, the first term corresponds to the wave packet contribution, originating from the zero point fluctuation, while the second term represents the edge contribution. In this example, the $Q$-tensor can be regarded as the quadrupole moment of the system, as originally introduced in Ref.~\cite{Daido2020}.

Now let us consider polarization responses described by Eq.~\eqref{eq:polarization_response}. The response coefficients $\chie$ and $\chit$ can be easily obtained from the $Q$-tensor using Eq.~\eqref{eq:Q-tensor_chie_chiT}. By identifying $\mu_0-e\Delta \phi$ with the local chemical potential $\mu(\vec{r})$ in the local equilibrium, we find
% obtain electric susceptibility $\chie$ and thermopolarizability $\chit$ as 
\begin{align}
	\chie^{\alpha\beta} &= -e\pdv{Q^{\alpha\beta}}{\mu}, \label{eq:polarizability_quad}\\
	\chit^{\alpha\beta} &= -\frac{1}{T} \pdv{Q^{\alpha\beta}}{T}. \label{eq:thermopolarizability}
\end{align}
Physically, an applied electric field or temperature gradient induces a spatial change in the $Q$-tensor, which in turn induces polarization through Eq.~\eqref{eq:P-Q}. 
At low temperatures $\kB T\ll E_g$ where $E_g$ is the band gap, the derivative of the first term in Eq.~\eqref{eq:quadrupole} becomes negligible, and Eq.~\eqref{eq:polarizability_quad} reduces to $\chi_0$ given by Eq.~\eqref{eq:polarizability}, consistent with discussion in Ref.~\cite{Daido2020}.

Eq.~\eqref{eq:polarizability_quad} parallels the Streda formula~\cite{streda1983}, which relates Hall conductivity to the $\mu$-derivative of magnetization. Here, $Q$-tensor plays the role of magnetization in the Streda formula. Notably, $\chie$ at low temperatures mostly comes from the ``edge'' contribution. In the edge picture, the polarization response is interpreted as the polarization difference at the two edges with a different local chemical potential.

From Eqs.~\eqref{eq:quadrupole} and \eqref{eq:thermopolarizability}, we can find a detailed expression for the thermopolarizability $\chit$. 
At low temperatures $\kB T\ll E_g$, neglecting the derivative of the first term in Eq.~\eqref{eq:quadrupole}, $\chit$ is: 
\begin{align}
	\chit^{\alpha\beta} = -eT \int[\dd{k}] \sum_n G_n^{\alpha\beta} s(\ene_n-\mu) \label{eq:thermopolarizability1}
\end{align}
where $s(\ene) = (\ene/T)f(\ene)+\kB\log[1+e^{-\beta\ene}]$ is the entropy density carried by the state at energy $\ene$~\cite{xiao2020a}. 
This result~\eqref{eq:thermopolarizability1} is consistent with the linear response theory (see Supporting Information for details), and also with general theory presented in Ref.~\cite{dong2020} by substituting the naive position operator $\vec{r}=i\nabla_{\vec{k}}$ into $\theta$ operator.

Although neglected in Eq.~\eqref{eq:thermopolarizability1}, the wave packet contribution from the first term in Eq.~\eqref{eq:quadrupole} is generally finite. This contribution arises from thermally activated carriers and is exponentially small at low temperatures. The ``edge'' contribution from the second term in Eq.~\eqref{eq:quadrupole} is also exponentially small, and both contribution can be of the same order. This work, however, focuses on the ``edge'' contribution in Eq.~\eqref{eq:quadrupole}, which reflects interband hybridization of quantum states, rather than the thermally activated carrier effect. 

Interestingly, the expression~\eqref{eq:thermopolarizability1} together with Eq.~\eqref{eq:polarizability} closely resembles other thermoelectric responses, such as the anomalous Nernst effect~\cite{Xiao2006}. 
$G_n$ here plays a role of the Berry curvature in the anomalous Hall and Nernst effects.
Similar to these effects, $\chit$ and $\chie$ satisfy the generalized Mott relation:
\begin{align}
	\chit(\mu, T) &= -\frac{1}{e} \int\dd{\ene} (\ene-\mu)\chie(\ene) \pdv{f(\ene-\mu)}{\mu} \label{eq:thermopolarizability2} 
\end{align} 
where $\chie(\mu) \equiv \chie(\mu, 0) = \int[dk] G_n^{\mu\nu} \Theta(\mu-\ene_n)$ is the electric susceptibility at $T=0$ for a chemical potential $\mu$. The generalized Mott relation was recently shown for the response of orbital magnetization and spin moment, and spin torque~\cite{freimuth2014, shitade2019, xiao2021a}. Here we extend the relation to the polarization response.

The origin of the thermopolarization effect can be understood with the edge picture. A temperature gradient causes a temperature difference at the two edges of the sample, resulting in a difference between the polarization at the edges and hence a net polarization, i.e., thermopolarization effect. Indeed, Eq.~\eqref{eq:thermopolarizability1} comes from the ``edge'' contribution to the $Q$-tensor, being consistent with the edge picture. We emphasize that the polarization at the edges does not require topological edge states, in contrast to the quantum Hall insulators. We also note that, while the edge picture is helpful for understanding the origin of thermopolarization effect, it remains a bulk property, as the $Q$-tensor is a bulk property.

The thermopolarization effect can contribute to the Seebeck effect~\cite{Marvan1969} in finite-size systems. The induced polarization generates an electric field $E=-P/\epsilon$ with $\epsilon = \epsilon_0 + \chie$ the dielectric constant of the system. Therefore the Seebeck coefficient $S$, defined as the ratio between the induced electric field $E$ and the applied temperature gradient $\nabla T$, is given by 
\begin{align}
	S = \frac{E}{\nabla T} = -\frac{\chit}{T\epsilon}.
\end{align}

\begin{figure}[tbp]
	\centering
	\includegraphics[width=0.8\columnwidth]{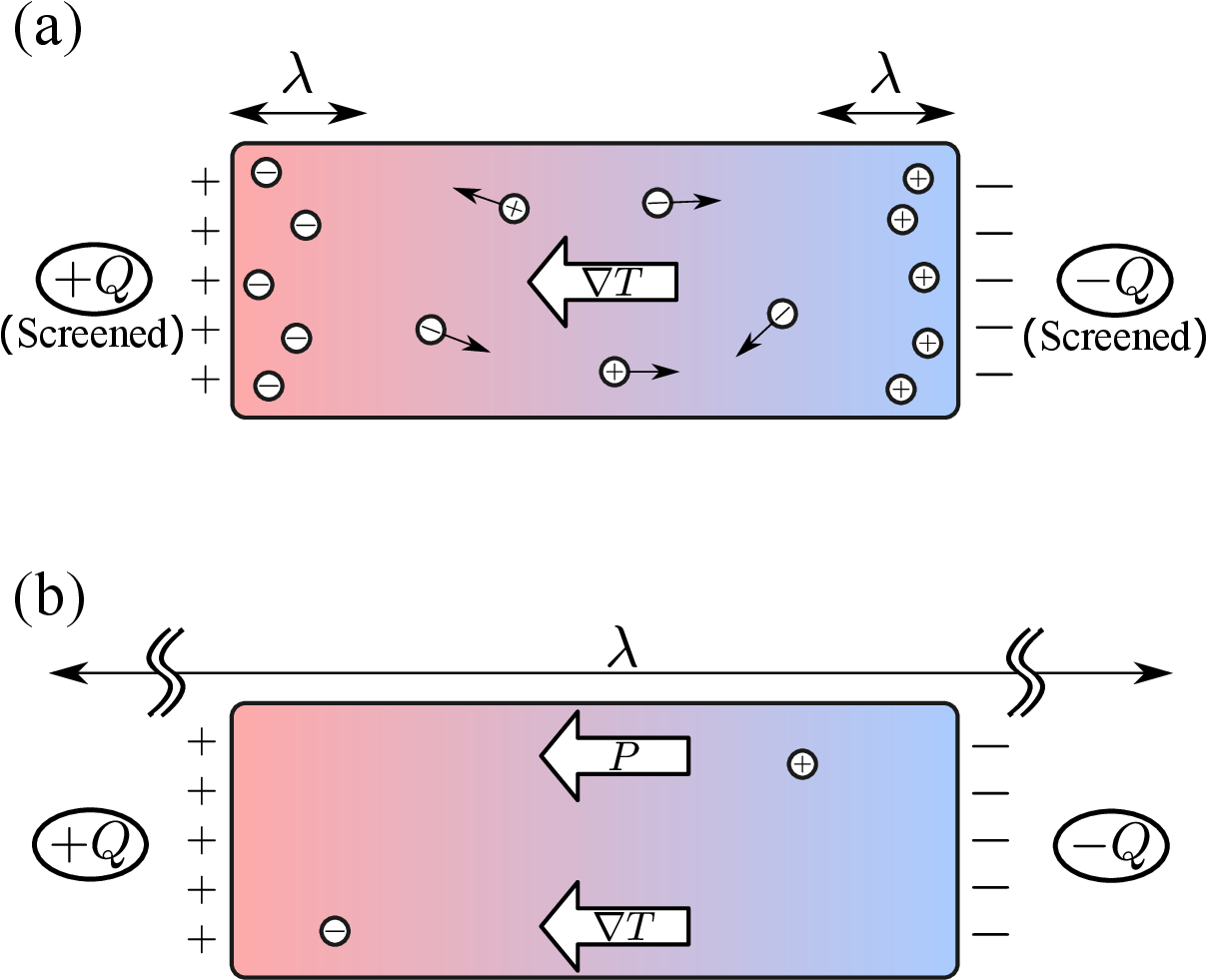}
	\caption{The thermopolarization effect and the screening due to finite carriers. (a) When the temperature is high and the system size $L$ is much larger than the screening length $\lambda$, $\lambda \ll L$, the polarization or the surface charge $\pm Q$ induced by the temperature gradient is screened and only the usual Seebeck effect, i.e., the one due to thermally activated carriers, can occur. (b) When the temperature is sufficiently low so that $\lambda \gg L$, the screening can be ignored and the polarization $P$ induced by the temperature gradient $\nabla T$ contributes to the Seebeck effect.}
	\label{fig:thermopolarization_screening}
\end{figure}

To observe the Seebeck effect from thermopolarization, it is crucial that the system size is finite so that the voltage is not screened by carrriers.
At finite temperatures, thermally excited carriers can screen the electric field induced by the polarization, thereby inhibiting the generation of a measurable voltage. This screening occurs when the screening length $\lambda$ is much smaller than the linear size of the system $L$ (Fig.~\ref{fig:thermopolarization_screening}(a)). %
Therefore, a finite voltage can be observed only when $L\ll \lambda$ (Fig.~\ref{fig:thermopolarization_screening}(b)). Since $\lambda$ is given by $\lambda = \sqrt{\epsilon\kB T/(ne^2)}$ with the electron density $n$, the condition is written as 
\begin{align}
	L\ll \lambda \sim \sqrt{\frac{\epsilon\kB T}{n_0e^2}} \exp(\frac{E_{c/v}-\mu}{2\kB T}). \label{eq:screening_condition}
\end{align}
Here, $n$ is assumed to be of the form $n=n_0 \exp(-(E_{c/v}-\mu)/(\kB T))$ where $E_{c/v}$ is either the energy of the conduction band or the valence band depending on the type of the carrier. Only when Eq.~\eqref{eq:screening_condition} is satisfied, the thermopolarization effect can be observed as the Seebeck effect. In this regime, the observed voltage is proportional to the linear size of the system $L$. 

Since the entropy density $s(\ene-\mu)$ is strongly localized near the chemical potential $\mu$ at low temperatures, the thermopolarizability $\chit$~\eqref{eq:thermopolarizability1} is exponentially small at low temperatures when $\mu$ lies within the gap, and scales as $\chit \propto \exp(-\beta (E_{c/v}-\mu))$. Therefore, the Seebeck coefficient $S$ scales as $S \sim (\kB/e) \exp(-(E_{c/v}-\mu)/(\kB T))$. This means that the thermopolarization effect can contribute to the Seebeck effect of at the same order as the usual contribution from the thermally activated carriers. The two contribution, polarization-induced and carrier-driven, can be distinguished by the sample size dependence of the system, as discussed above. 

\begin{figure}[htbp]
	\centering
	\includegraphics[width=0.8\columnwidth]{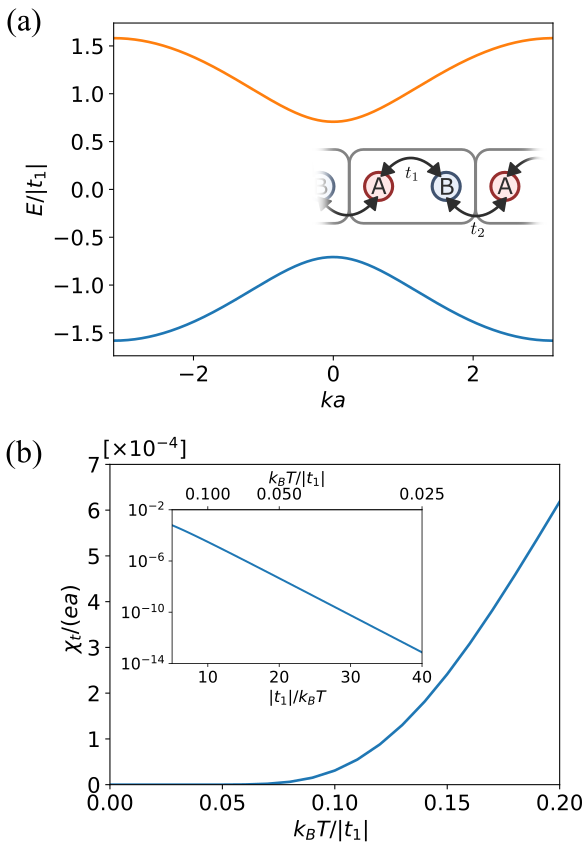}
	\caption{Thermopolarizability of SSH model~\eqref{eq:SSH} with $t_{1}=-1.0$, $t_{2}=0.5$, and $\Delta=1.0$. $r_A=0.0, r_B=0.5a$ with a lattice constant $a$, and the chemical potential is set to be $\mu=0.05\abs{t_1}$. (a) The band structure of the model. The inset is a schematic figure of the SSH model. (b) The thermopolarizability as a function of temperature $T$. The inset shows the exponential dependence of the thermopolarizability at low temperatures. See Supporting Information for more details of the calculation. }
	\label{fig:SSH}
\end{figure}

To illustrate the thermopolarization effect, we consider a simple model of a one dimensional Su-Schrieffer-Heeger model described by the following Hamiltonian:
\begin{align}
	H = \sum_n (t_{1} \cre{c}_{nB} \ani{c}_{nA} + t_{2} \cre{c}_{n+1,A} \ani{c}_{nB} + \mathrm{h.c.}) + \frac{\Delta}{2} (\cre{c}_{nA} \ani{c}_{nA} - \cre{c}_{nB} \ani{c}_{nB}) \label{eq:SSH}
\end{align}
where $\ani{c}_{n{\alpha}}$ is an annihilation operator for site $\alpha=A,B$ in $n$-th unit cell, each localted at $na+r_{\alpha}$ with lattice constant $a$. Site $A$ and $B$ in the same unit cell are connected by the hopping $t_{1}$, and the hopping between sites $A$ and $B$ in neighboring unit cells is $t_{2}$. The onsite energy difference between the sites $A$ and $B$ is $\Delta$. 
We calculated the thermopolarizability $\chit$ for this model with Eq.~\eqref{eq:thermopolarizability1}, and the results are shown in Fig.~\ref{fig:SSH}. 
Although the effect is exponentially small, a finite thermopolarizability persists at low temperatures.

The ``dual'' effect to the thermopolarization effect is the heat current generation by the AC electric field, which is proportional to $\dot{E}$. Indeed, the thermopolarization effect can be viewed as the AC electric current response to an AC temperature gradient: $\vec{J}=-i\omega \vec{P}=-i\omega \chit (-\nabla T/T)$ with frequency $\omega$. Therefore, the dual effect to the thermopolarization effect is $\vec{J}^Q = -i\omega \tilde{\chit} \vec{E}$, or equivalently, 
\begin{align}
	\vec{J}^Q = \tilde{\chit} \dot{\vec{E}}. \label{eq:ac_heat_current}
\end{align}
The response coefficient of this effect, $\tilde{\chit}$, should be equal to $\chit$ under the time reversal symmetry because of the Onsager reciprocity: $\tilde{\chit}=\chit$. Note that the static electric field $\vec{E}$ cannot induce a heat current in insulators because it only shifts the system to a new equilibrium state without any current flow.
This asymmetry between the DC polarization and AC heat current is analogous to the Edelstein effect in semiconductors~\cite{Edelstein1990}, where magnetization $M$ and charge current $J$ are given by
\begin{align}
& M  = \chi_M H + \chi_{ME} E,  
\\ 
& J = \chi_{ME} \dot{H} + \sigma E, 
\label{eq:linear4}
\end{align}
where $H$ is the magnetic field and $E$ is the electric field.
Here, the magnetic field simply shifts the equilibrium state to a magnetized one, while $E$ drives the system into a nonequilibrium state with a charge current. The heat current response to AC electric field~\eqref{eq:ac_heat_current} may provide another way to observe the thermopolarization effect.

Our calculation is fully based on quantum theory of electrons, and thus our results on thermopolarization effect show that the polarization, arising from quantum hybridization of electronic states, can be induced by a statistical force. This contrasts with earlier studies on the thermopolarization effect, which attributed the polarization to ionic position shift~\cite{Tagantsev1987,Marvan1969} or charge redistribution~\cite{Tagantsev1991}.  

Before concluding, we comment on the consistency with the linear response theory. The calculation for thermopolarizability $\chit$ with simple Kubo formula does not reproduce Eq.~\eqref{eq:thermopolarizability1}, and the results will diverge as $T\to 0$. This situation is similar to the calculation of the anomalous Nernst effect, where the correction to the Kubo formula due to the orbital magnetization is necessary to remedy unphysical divergence at $T\to0$~\cite{Qin2011}. 
To correctly derive $\chit$ from the linear response theory, we need a correction involving the $Q$-tensor. With this correction, the thermopolarizability $\chit$ converges to zero at $T=0$ and coincides with the results from the thermodynamic discussion presented above. For more details, see Supporting Information and references therein~\cite{Kubo1957,Kubo1957a,Mori1958,Luttinger1964,shitade2014,Tatara2015,Shastry2009,Qin2011,Mahan2000,zhang2020b,Watanabe2021}.

In summary, we discussed polarization responses based on thermodynamic $Q$-tensors in a unified manner, including the thermopolarization effect.
The discussed effects are purely due to quantum mechanical electrons in contrast to the thermopolarization effect investigated in previous studies~\cite{Tagantsev1987,Tagantsev1991,Marvan1969}. The thermopolarization effect contributes to the Seebeck effect in finite-size system at low enough temperatures satisfying Eq.~\eqref{eq:screening_condition} so that the system size is shorter than the screening length.

% \begin{acknowledgement}
	\acknowledgements
	The authors are grateful to S. Maekawa, B. S. Shastry, A. Daido, Y. Yanase, Y. Gao, and D. Xiao for insightful discussions and useful comments. 
	N.N. was supported by JSPS KAKENHI Grant Numbers JP24H00197 and JP24H02231. N.N. was supported by the RIKEN TRIP initiative. A.S was supported by JSPS KAKENHI Grant Numbers JP21H01816 and JP22K03498.
% \end{acknowledgement}

\nocite{Mahan2000,Qin2011,Shastry2009}
% \bibliography{library_zotero, arxiv}
\bibliography{thermopolarization, arxiv}
% \bibliography{library_zotero}

\newpage

\clearpage

\newpage

\appendix
\begin{widetext}
\section*{Supporting Information}
\section{Linear response theory of thermopolarization effect}
As is well known, treatment of polarization in periodic systems is theoretically a subtle problem because the position operator is not well-defined in periodic systems. We can avoid this difficulty by considering a response with a finite frequency $\omega$. Since the time derivative of the polarization $P$ is equal to the current in insulators, $\vec{J}=-i\omega\vec{P}$ holds and we can extract the polarization response from the current response. 
% Here, we calculate the polarization response from the current response at finite frequency by using the relation between the polarization and the current. 

% This difficulty can be avoided by considering the change of polarization instead of polarization itself. Indeed, only the change of polarization for a certain process is a gauge invariant quantity, and it is related to the underlying topology of the process. 

% In order to calculate the polarization response, we need to consider the change of the polarization during an adiabatic application of an external field. For example, the response to an electric field, i.e., polarizability $\chie$ can be obtained as the change of the polarization during an adiabatic application of an external electric field. This calculation can be done by introducing adiabatic switching of the electric field as $\vec{E}(t) \propto e^{\eta t}$, or, equivalently, considering response with finite frequency $\omega$ and calculating the current response proportional to $\dv*{\vec{E}}{t} \propto \omega \vec{E}$. The polarization is obtained as the integrated current during the adiabatic response, $\Delta P=\int \dd{t}\vec{j}(t)$. Similarly, we need to consider finite frequency response to a temperature gradient $\grad{T}\propto e^{-i\omega t}$ and integrate the electric current proportional to $\dv*{\grad{T}}\propto \omega \grad{T}$ to calculate the thermopolarization response. 

To calculate the finite frequency response, we will use the linear response theory. The linear response theory can be formulated in terms of the correlation functions of charge current and heat current. 
Thermal linear response requires additional consideration, since the temperature gradient $\nabla T$ is not described by the Hamiltonian~\cite{Kubo1957, Kubo1957a, Mori1958}.
This is in sharp contrast to the case of electromagnetic field. 
% where the charge current density can be defined as the functional derivative of the action with respect to the vector potential. 
Luttinger~\cite{Luttinger1964} proposed a method to formulate the thermal response by introducing the gravitational field $\psi(\bm{r},t)$, which couples to the energy density.
%  is defined as the functional derivative of the action with respect to $\psi(\bm{r},t)$. 
Then, the response of the system to $\nabla \psi(\bm{r},t)$ replaces that of $\nabla T$. A heat analogue of the vector potential which couples to the energy current was also proposed to calculate the thermal responses~\cite{shitade2014, Tatara2015}.
% Tatara~\cite{Tatara2015} introduced the thermal vector potential $\bm{A}_T(\bm{r},t)$ in addition to $\psi(\bm{r},t)$, and defined the heat current density $\bm{J}^Q(\bm{r},t)$ as the functional derivative with respect to $\bm{A}_T(\bm{r},t)$. 
More explicitly, the linear response theory is formulated as~\cite{Shastry2009}
\begin{align}
& J_x  = L_{11} E_x + L_{12}( - \nabla_x T/T) + \Lg_{12} ( - \nabla_x \psi ), \label{eq:linear_J}\\ 
& J^Q_{x} = L_{21} E_x + L_{22}( - \nabla_x T/T) + \Lg_{22} ( - \nabla_x \psi ), 
\label{eq:linear_JQ}
\end{align}
where $J_x$ ($J^Q_x$) is the $x$-component of
the charge (heat) current, and $L_{ij}$ and $\Lg_{ij}$ are the linear response coefficients.
$\Lg_{12}$ and $\Lg_{22}$ can be calculated as the mechanical response to the gravitational potential $\psi$ while $L_{12}$ and $L_{22}$ cannot be described in a similar way because there is no corresponding potential in the Hamiltonian. 
Luttinger argued that $L_{ij}$ and $\Lg_{ij}$ are identical in the limit of $\bm{q} \to 0$ and $\omega =0$ where $\bm{q}$ is the wavevector and $\omega$ is the frequency of the response functions because of a generalized Einstein relation. 
Shastry~\cite{Shastry2009} extended this argument to the finite frequency case, i.e., proposed 
\begin{align}
L_{ij}( \bm{q}, \omega) = \Lg_{ij}( \bm{q}, \omega), 
\label{eq:Shastry}
\end{align}
for generic $\vec{q}$ and $\omega$. This idea seems natural at least for small $\vec{q}$ and $\omega$, and we follow it to consider the thermal response at finite frequencies, which enables us to define the thermal response of insulating systems, and identify $\Lg$ with $L$ hereafter.
If the system has the time-reversal symmetry, the Onsager theorem~\cite{Shastry2009} indicates $L_{ij}(\bm{q}, \omega) = L_{ji}( - \bm{q}, \omega)$.
% , leading to $\Pi(\omega)= L_{21}(\bm{0}, \omega)/L_{11}(\bm{0}, \omega)$ and $S(\omega) = \Pi(\omega)/T$. 
% In this formula applied to insulators at $\omega=0$, we can show that both $L_{12}(=L_{21})$ and $L_{11}$ are due to the thermally excited electrons with the density $n \sim e^{- E_G/2\kB T}$ and thus $L_{12}$ and $L_{11}$ vanish as $T \to 0$. 
% However, approaching from finite $\omega$, one can define 
% $\Pi(\omega)$ even at $T=0$ for insulators since both 
% $L_{11}(\omega)$ and $L_{21}(\omega)$ are proportional to
% $\omega$ as $\omega \to 0$.   
% This means that the two limits, i.e., $T \to 0$ and $\omega \to 0$ cannot be exchanged, and the order of these two limits is essential. To make this point clear, it is important to consider an adiabatic factor $\eta$, which is introduced below. 
Since the polarization $P$ is related to the electric current $J$ as $\vec{J}=-i\omega\vec{P}$, the polarization response corresponds to the $\omega$-linear term in the current response. Therefore, the DC electric susceptibility $\chi_e$ is given by the $\omega$-linear term in $L_{11}$ at small $\omega$, while the thermopolarizability is given by the $\omega$-linear term in $L_{12}$ (or $L_{21}$).

We should note that, there is sometimes an additional contribution to Kubo formula to obtain physically reasonable results. For example, when the time reversal symmetry is broken, there will be a contribution from magnetization and energy magnetization to the Kubo formula for $L_{ij}$ at $\omega=0$~\cite{Qin2011}. 
Similarly, there will be two contributions to $\chit$:
\begin{align}
	\chit &= \chit^{\mathrm{Kubo}} + \chit^{\mathrm{Q}}.
\end{align}
The first term $\chit^{\mathrm{Kubo}}$ is the contribution calculated from the Kubo formula. The second term $\chit^{\mathrm{Q}}$ is a contribution analogous to the magnetization current, and we shall call it "$Q$-tensor contribution" because of the reason which will be clear later.

The $Q$-tensor contribution $\chit^{\mathrm{Q}}$ can be further divided into two contributions: surface contribution and quantum contribution. The surface contribution is given by the polarization induced at the surface, while the quantum contribution is given by so-called quantum metric.

\subsection{Calculation of Kubo formula contribution} \label{sec:calc_liners}

In this section, we calculate the linear response coefficients from the Kubo formula calculation, $L_{ij}^{\mathrm{Kubo}}(\omega)$. We do not consider the vertex correction. 
% Since we are interested in longitudinal responses of the system with the time-reversal symmetry, we do not need modification due to magnetization or energy magnetization~\cite{Qin2011}. 
Therefore, $L_{ij}^{\mathrm{Kubo}}(\omega)$ can be obtained from current-current correlation functions as follows~\cite{Mahan2000, Shastry2009} (we set $\hbar=1$):
\begin{align}
	L^{\rm Kubo}_{ij}(\omega) &= \frac{\Phi^R_{ij}(\omega)-\Phi^R_{ij}(0)}{i(\omega+i\eta)}, \label{eq:response_function}\\
	\Phi^R_{ij}(\omega) &= \eval{\Phi_{ij}(i\omega_\lambda)}_{i\omega_\lambda \to \omega + i\eta}, \\
	\Phi_{ij}(i\omega_\lambda) &= \frac{1}{V} \int_0^\beta \dd{\tau} \expval{\tauord \hat{J}_i(\tau) \hat{J}_j} e^{i\omega_\lambda \tau}, \label{eq:corr_func}
\end{align}
where $i,j = 1,2$ and $\hat{J}_1 = (\hat{\vec{J}})_x, \hat{J}_2 = (\hat{\vec{J}}^Q)_x$. $\omega_\lambda = 2\pi\lambda/\beta$ is the bosonic Matsubara frequency,  and $V$ is the volume of the system. $\expval{\dots}$ is the expectation value in the equilibrium and $\tauord$ is the ordering operator for imaginary time $\tau$. $\beta=1/\kB T$ is an inverse temperature, $\tau$ is an imaginary time and $\hat{\mathcal{O}}(\tau) = e^{\tau\hat{H}}\hat{\mathcal{O}}e^{-\tau\hat{H}}$ is the Heisenberg representation of an operator $\hat{\mathcal{O}}$. $\eta$ is an infinitesimal positive quantity in the thermodynamic limit but finite in finite size systems, as explained at the end of this section.

We can evaluate the correlation function Eq.~\eqref{eq:corr_func} with Matsubara Green's functions $\G(i\ene_n, \vec{k})$:
\begin{align}
	(\G(i\ene_n, \vec{k}))_{l,l'} &= \int_0^\beta \dd{\tau} e^{i\ene_n \tau}\expval{\tauord \ani{d}_{\vec{k},l}(\tau)\cre{d}_{\vec{k},l'}} = \delta_{l,l'} \G_{l}(i\ene_n, \vec{k}), \\
	\G_{l}(i\ene_n, \vec{k}) &= \frac{1}{i\ene_n-\xi_{\vec{k},l} + i\Gamma_{\vec{k},l}\mathrm{sgn}(\ene_n)}. 
\end{align}
Here we have introduced $\Gamma_{\vec{k},l}$ which is related to the lifetime of the electron of the $l$-th band at wavevector $\vec{k}$ as $\tau_{\vec{k},l}=1/(2\Gamma_{\vec{k},l})$, and $\xi_{\vec{k},l}=\ene_{\vec{k},l}-\mu$. $\ene_n=(2n+1)\pi/\beta$ is the fermionic Matsubara frequency. Then the correlation function Eq.~\eqref{eq:corr_func} reads
\begin{align}
	\Phi_{ij}(i\omega_\lambda) %\nonumber\\
	&= -\frac{\kB T}{V}\sum_{\vec{k}, n}\sum_{l}(J_{i, \vec{k}})_{l,l}\G_l(i\ene_n, \vec{k})(J_{j, \vec{k}})_{l,l} \G_l(i\ene_n-i\omega_\lambda, \vec{k}) \nonumber\\
	&-\frac{\kB T}{V}\sum_{\vec{k}, n} \sum_{l\neq l'}(J_{i, \vec{k}})_{l,l'}\G_{l'}(i\ene_n, \vec{k})(J_{j, \vec{k}})_{l',l} \G_l(i\ene_n-i\omega_\lambda, \vec{k}). \label{eq:corr_func_green}
\end{align}
The first term represents the intraband contribution and the second term represents the interband contribution.

This correlation function can be evaluated with the common technique of the Matsubara summation (see Appendix~\ref{Ap:calc_linres} for details). After the analytic continuation, we obtain the results for $L_{11}$ and $L_{12}$ without vertex corrections up to $\order{\omega}$:
\begin{align}
	L_{11}^{\mathrm{Kubo}}(\omega) &= \sigma -i\omega \chie,  \label{eq:L11_Kubo} \\
	L_{12}^{\mathrm{Kubo}}(\omega) &= \alpha - i\omega \chit^{\mathrm{Kubo}}. \label{eq:L12_Kubo}% + \order{\omega^2}
\end{align}
Here, $\sigma, \chie, \alpha, \chit$ are given by 
\begin{align}
	\sigma &= -\frac{e^2}{V}\sum_{\vec{k},l}f'(\ene_{\vec{k},l}-\mu)(v_{l,\vec{k},x})^2 \tau_{\vec{k},l}(1+i\omega\tau_{\vec{k},l}), \\
	\chie &= -\frac{2e^2}{V} \sum_{\vec{k}, l\neq l'} \frac{\abs{(A_{\vec{k},x})_{ll'}}^2}{\ene_{\vec{k},l} - \ene_{\vec{k},l'}} f(\ene_{\vec{k},l}-\mu), \\
	\alpha &= -\frac{e}{V}\sum_{\vec{k},l}f'(\ene_{\vec{k},l}-\mu)\qty(\ene_{\vec{k},l}-\mu)(v_{l,\vec{k},x})^2 \tau_{\vec{k},l}(1+i\omega\tau_{\vec{k},l}), \label{eq:alpha}\\
	\chit^{\mathrm{Kubo}} &= -\frac{2e}{V} \sum_{\vec{k}, l\neq l'} \frac{\abs{(A_{\vec{k},x})_{ll'}}^2}{\ene_{\vec{k},l} - \ene_{\vec{k},l'}} \qty(\frac{\ene_{\vec{k},l}+\ene_{l',\vec{k}}}{2}-\mu)f(\ene_{\vec{k},l}-\mu), \label{eq:L12_tilde}
\end{align}
where $f(\ene)=(e^{\beta\ene}+1)^{-1}$ is the Fermi distribution function with $\beta=1/(\kB T)$, $v_{l,\vec{k},x}=\partial{\ene_{\vec{k},l}}/\partial{k_x}$ is the group velocity and $\tau_{\vec{k},l}=1/(2\Gamma_{\vec{k},l})$.
From Eq.~\eqref{eq:alpha}, we can perform an order-of-magnitude estimate of $\alpha$. Considering only one band and assuming that the dispersion is parabolic and the lifetime $\tau_{\vec{k}}$ is independent of $\vec{k}$, then $\sigma$ becomes the common expression $\sigma = ne^2\tau/m$ for $\omega = 0$. If we estimate $\ene_{\vec{k},l}-\mu$ as the band gap $E_G$, then $\alpha\sim E_G\sigma/e\sim neE_G\tau/m$. 

Notably, the above results based on the Kubo formula for the thermopolarizability is different from the results we obtained with thermodynamic arguments in the main text. In particular, $\chit^{\mathrm{Kubo}}$ does not vanish in $T\to0$ limit and also depends on the chemical potential $\mu$ even at zero temperature. Such physically unreasonable behaviors can be remedied by considering the $Q$-tensor contribution $\chit^{\mathrm{Q}}$ so that we can obtain a consistent result with the main text, as we will see in the next section.

\subsection{$Q$-tensor contribution}
The Kubo formula calculation in the previous section gives the bulk contribution to the polarization or the electric current $\vec{j}$. However, its result has two problems. First, it does not take into account any surface contribution. In real materials, the system always has a surface, which can be important in polarization calculation. The other problem is the contribution from the bulk, but from the correction to the current operator due to external fields. This correction is related to the quantum metric and we shall call it the quantum contribution. These two contributions are discussed in this section, and we shall call them together as the $Q$-tensor contribution because of the reason which will be clear later.

As the Onsager relation shows, the thermopolarizability $\chit$ also represents the heat current response to the electric field at finite frequency: $J^Q_{x} = (\alpha-i\omega \chit) E_x$. In the following, we consider the heat response to the electric field, instead of the electric response to the temperature gradient. For simplicity, we consider one-dimensional systems.
% Equivalently, we can consider the response of ``heat'' polarization $P^Q$ to the electric field. 
% For finite size systems, $P^Q$ can be defined as $P^Q = \int\dd{r} \expval{\vec{r}(\hat{h}(\vec{r})-\mu\hat{\rho}(\vec{r}))}$ and as in the case of polarization, $\chit$ should represent the response of $P^Q$ to the electric field:
% % \begin{align}
% % 	J^Q_{x} = (\alpha-i\omega \chit) E_x. 
% % \end{align}
% % By integrating over time, the response corresponding to $\chit$ can be also considered as a response of {\it energy polarization} $P_E = $ to electric field, namely, 
% \begin{align}
% 	P_Q = \chit E_x.
% \end{align} 
% In the following, we consider the response of $P_Q$. 

Let us first consider the contribution from the surface. At surface, there exists a confining potential $\phi_c(x)$ 
% The surface contribution can be taken into account by introducing a confining potential $\phi_c(x)$, 
as considered in the calculation of orbital magnetization in literature~\cite{zhang2020b}. $\phi_c(x)$ contributes to the energy around the surface and thus contributes to the heat current response. 

$\phi_c(x)$ is assumed to be zero in the bulk ($x<0$) and increase as $x$ increases so that the density $\expval{\hat{\rho}(x)}=0$ at $x=\infty$. Due to the continuity relation of heat, the heat current flowing into the surface ($x>0$) should be given by the energy change of the surface measured from the chemical potential, namely, 
\begin{align}
	\expval{j^Q(x=0)} &= \dv{E_{\rm surf}}{t} 
	% = \dv{}{t}\int_0^d\dd{x}\expval{(\hat{h}(x)-\mu\hat{\rho}(x))} %\\
	% \nonumber \\
	= \dv{}{t}\int_0^\infty\dd{x} \expval{\hat{h_0}(x) - \mu\hat{\rho}(x) + \phi_c(x)\hat{\rho}(x)} \nonumber \\ 
	% = -\int_0^\infty\dd{x}\qty(\pdv{\expval{j^Q_0}}{x}+\phi_c(x)\pdv{\expval{\hat{j_x}}}{x}) \nonumber \\
	&= \expval{j^Q_0(x=0)} + \expval{j^Q_{\rm surf}(x=0)}. \label{eq:edge_heat_polarization}
\end{align} 
where $\hat{h}_0(x)$ is the Hamiltonian density without the confining potential, $\hat{\rho}(x)$ is the electron density and
% Let us rewrite $P^Q$ with the heat current. Using the continuity equation on the heat and charge, the time derivative of $P^Q$ is given by
% \begin{align}
% 	\dv{P^Q}{t} &= \int_0^\infty\dd{x} \expval{\pdv{(\hat{h}_0(x)-\mu\hat{\rho}(x))}{t} + \pdv{\hat{\rho}(x)}{t}\phi_c(x)} \nonumber \\
% 	&= \int_0^d\dd{x}\qty(-\pdv{\expval{j^Q_0}}{x}-\pdv{\expval{\hat{j_x}}}{x}\phi_c(x)) \nonumber \\
% 	&= \expval{j^Q_0(x=0)} + \int_0^\infty\dd{x} \expval{\hat{j_x}(x)}\pdv{\phi_c(x)}{x}. \label{eq:edge_heat_polarization}
% \end{align}
$j^Q_0(x)$ is the heat current operator when there is no confining potential. The first term is the bulk contribution and should be described by the Kubo formula. The second term is the surface contribution we would like to discuss here, given by 
% On the other hand, due to the continuity relation, $\dv*{E_{\rm surf}}{t}=\expval{j^Q(x=0)}$ hold, and thus the second term in Eq.~\eqref{eq:edge_heat_polarization} gives the surface contribution to the heat current response $\expval{j^Q_{\rm surf}(x=0)}$:
\begin{align}
	% = -\int_0^\infty\dd{x}\qty(\pdv{\expval{j^Q_0}}{x}+\phi_c(x)\pdv{\expval{\hat{j_x}}}{x}) \nonumber \\
	\expval{j^Q_{\rm surf}(x=0)} = -\int_0^\infty\dd{x} \phi_c(x)\pdv{\expval{\hat{j_x}(x)}}{x}. \label{eq:edge_heat_polarization2}
\end{align}
We can further rewrite $\expval{\hat{j}^Q_{\rm surf}(x=0)}$ in an analogous way to Ref.~\cite{zhang2020b}. Assuming $\phi_c(x)$ is sufficiently slowly varying and free carrier contribution is negligible, we can express the current at $x$ as $\expval{\hat{j_x}(x)}=-i\omega\chie(\mu-e\phi_c(x))E_x$ with $\chie(\mu)$ the electric susceptibility at chemical potential $\mu$ and $E_x$ the applied electric field with frequency $\omega$. Then the second term in Eq.~\eqref{eq:edge_heat_polarization} becomes
\begin{align}
	\expval{j^Q_{\rm surf}(x=0)} &= \int_0^\infty\dd{x} \expval{\hat{j_x}(x)}\pdv{\phi_c(x)}{x}= \int_0^\infty\dd{x} \qty(-i\omega\chie(\mu-e\phi_c(x))E_x)\pdv{\phi_c(x)}{x} \nonumber \\
	% &= -i\omega \int_0^\infty\dd{\phi_c}  \qty(\chie(\mu-\phi_c)E_x) \nonumber \\
	&= -i\omega \frac{1}{e} \int_{-\infty}^\mu\dd{\mu} \chie(\mu)E_x.
\end{align}
Therefore, the heat current contribution from the surface proportional to $\omega$ is given by the integral of the polarization over the chemical potential, or equivalently, the surface contribution to $\chit$ is given by 
\begin{align}
	\chit^{\mathrm{surf}} = \frac{1}{e}\int_{-\infty}^\mu\dd{\mu} \chie(\mu) 
\end{align}

Another contribution to the Kubo formula comes from the correction to the current operators due to the external fields. Let us consider the heat polarization response. The Bloch Hamiltonian under the electric field is given by 
\begin{align}
	\hat{H}(\vec{k}, t) = \hat{H}_0(\vec{k}) - e\vec{E}(t)\cdot\hat{\vec{r}}_{\vec{k}},
\end{align}
where $\hat{H}_0(\vec{k})$ is the Bloch Hamiltonian in the absence of external fields and the second term represents the applied electric field. The position operator $\vec{r}_{\vec{k}}=(x_{\vec{k}}, y_{\vec{k}}, \dots)$ here is defined in the band basis as $(\vec{r}_{\vec{k}})_{nm} = i\nabla_{\vec{k}} \delta_{nm} + \vec{A}_{nm}(\vec{k})$, where $\vec{A}_{nm}(\vec{k}) = \bra{n,\vec{k}}i\nabla_{\vec{k}}\ket{m,\vec{k}}$ is the interband Berry connection. While in general the position operator in the periodic system is not well-defined, it is known that the position operator defined in this way will give the consistent result for the conductivity in the linear response and the nonlinear responses with other calculation results~\cite{Watanabe2021}. 
% Therefore, the heat current operator in the presence of external fields is given by 

The heat current operator is given by 
\begin{align}
	\hat{j}^Q_{x}=\int[\dd{k}]\frac{1}{2\hbar}\qty(H\pdv{H}{k_x}+\pdv{H}{k_x}H)
	% =\int[\dd{k}]\frac{-i}{2\hbar^2}(H\comm*{x_{\vec{k}}}{H}+\comm*{x_{\vec{k}}}{H}H)
	=\int[\dd{k}]\frac{-i}{2\hbar}\comm*{Hx_{\vec{k}}+x_{\vec{k}}H}{H}
	=\int[\dd{k}]\frac{1}{2}\dv{t}(Hx_{\vec{k}}+x_{\vec{k}}H)
\end{align}
and thus the correction due to the electric field in the linear order, denoted as $\hat{j}^Q_{x,1}$, is given by 
\begin{align}
	\hat{j}^Q_{x,1} = -e\dv{}{t}\qty(E_x(t)\int[\dd{k}]\qty(\hat{x}_{\vec{k}})^2).
\end{align}
with $[dk]=d^dk/(2\pi)^d$. Therefore, the $\omega$-linear contribution to the heat current is given by the expectation value of $\hat{x}_{\vec{k}}^2$. Since the expectation value of the diagonal component of $\hat{x}_{\vec{k}}$ vanishes (note $\nabla_{\vec{k}}$ in $\vec{r}_{\vec{k}}$ acts on the coefficient when a state is expanded with $\ket{u_{n\vec{k}}}$), the expectation value of $\qty(\hat{x}_{\vec{k}})^2$ is given by $\expval{\qty(\hat{x}_{\vec{k}})^2} = \sum_{n} f(\ene_n-\mu) g_n^{xx}$ with the quantum metric $g_n^{\alpha\beta}=\sum_{m(\neq n)}\Re[A^\alpha_{nm}A^\beta_{mn}]$ and $f(\ene)$ the Fermi distribution function. The resulting correction to $\chit$ is given by
\begin{align}
	\chit^{\rm quantum} &= -e\int[\dd{k}]\sum_{n} f(\ene_n-\mu) g_n^{xx}.
\end{align}
This is the quantum contribution to the thermopolarizability.

The total correction to $\chit$ is given by the sum of the surface contribution and the quantum contribution:
\begin{align}
	\chit^{Q} &= \chit^{\rm surf} + \chit^{\rm quantum} 
	= -e\sum_n \int[\dd{k}] g_n^{\alpha\beta} f(\ene_n-\mu) + \frac{1}{e} \int_{-\infty}^\mu\dd{\mu'} \chie^{\alpha\beta}(\mu') 
	= -Q^{xx},
\end{align}
where $Q^{xx}$ is the $Q$-tensor moment. Namely, the correction to the Kubo formula for the thermopolarizability is given by the $Q$-tensor moment. This is similar to the correction to the Kubo formula for the anomalous Nernst effect where the orbital magnetization appears~\cite{Qin2011}.
The thermopolarizability $\chit$ is given by the sum of $\chit^{\rm Kubo}$ and $\chit^{Q}$:
\begin{align}
	\chit = \chit^{\rm Kubo} + \chit^{Q} = -\frac{1}{e} \sum_n \int[\dd{k}] G_n^{xx} \qty{(\ene_n-\mu)f(\ene_n-\mu) + \kB T \log[1+e^{-\beta(\ene_n-\mu)}]}.
\end{align}
This coincides with the result obtained from the thermodynamic argument in the main text.

\section{Current operators in noninteracting tight-binding models}
Here we derive the current operators in general noninteracting tight-binding models:
\begin{align}
	\hat{H} &= \sum_{\vec{R},\vec{R}'}\cre{c}_{\vec{R}} H_{\vec{R}-\vec{R}'}\ani{c}_{\vec{R}'} = \sum_{\vec{k}} \cre{c}_{\vec{k}} H_{\vec{k}} \ani{c}_{\vec{k}}, \label{eq:Hamiltonian}\\
	% \cre{c}_{\vec{R}} &= \qty(\cre{c}_{\vec{R},1}, \cre{c}_{\vec{R},2}, \dots, \cre{c}_{\vec{R},s})
	(H_{\vec{k}})_{ij} &= \sum_{\vec{R}} e^{-i\vec{k}\vdot(\vec{R}+\vec{r}_i-\vec{r}_j)} (H_{\vec{R}})_{ij}, \\
	\ani{c}_{\vec{R}} &= \qty(\ani{c}_{\vec{R},1}, \ani{c}_{\vec{R},2}, \dots, \ani{c}_{\vec{R},s})^T, \\
	\ani{{c}}_{\vec{k},i} &= \sum_{\vec{R}} \frac{1}{\sqrt{N}}e^{-i\vec{k}\vdot(\vec{R}+\vec{r}_i)} \ani{c}_{\vec{R},i}.
\end{align}	
Here, $\vec{R}$ specifies the position of a unit cell. $\ani{c}_{\vec{R}, i}$($\cre{c}_{\vec{R}, i}$) is an annihilation (creation) operator, which annihilates(creates) an electron in the $i$-th orbital of the unit cell at $\vec{R}$. $\ani{c}_{\vec{k}, i}$($\cre{c}_{\vec{k},i}$) is the Fourier transformation of $\ani{c}_{\vec{R}, i}$($\cre{c}_{\vec{R}, i}$). $N$ is the total number of unit cells and $\vec{R}+\vec{r}_i$ is the position of the $i$-th orbital in the unit cell at $\vec{R}$. $H_{\vec{R}-\vec{R}'}$ and $H_{\vec{k}}$ are $s\times s$ matrices. 

% The Hamiltonian in the Fourier representation is 
% \begin{align}
% 	\hat{H} &= \sum_{\vec{k}} \cre{c}_{\vec{k}} H_{\vec{k}} \ani{c}_{\vec{k}} \\
% 	(H_{\vec{k}})_{ij} &= \sum_{\vec{R}} e^{-i\vec{k}\vdot(\vec{R}+\vec{r}_i-\vec{r}_j)} (H_{\vec{R}})_{ij}
% \end{align}

% \subsection{Energy current operator and heat current operator}
To calculate the thermoelectric responses of the model, we need the energy current operator and the heat current operator.
To obtain the total energy current operator $\hat{\vec{J}}^E$, we follow the method in~\cite{Mahan2000}. We use the following operator $\hat{\vec{R}}_E$, which we call here an energy polarization operator:
\begin{align}
	\hat{\vec{R}}_E &= \sum_{\vec{R},i} \hat{h}_i(\vec{R}) (\vec{R} + \vec{r}_i), \\
	\hat{h}_i(\vec{R}) &= \frac{1}{2}\sum_{\vec{R}'}\qty(\cre{c}_{\vec{R}}P_i H_{\vec{R}-\vec{R}'} \ani{c}_{\vec{R}'} + \cre{c}_{\vec{R}'}H_{\vec{R}'-\vec{R}}P_i \ani{c}_{\vec{R}}),
\end{align}
where $\hat{h}_i(\vec{R})$ is the Hamiltonian density at $i$-th orbital in the unit cell $\vec{R}$. $P_i$ is a matrix which projects on $i$-th orbital, and its matrix element is given by   
% \begin{align}
$(P_i)_{jk} = \delta_{ij}\delta_{ik}$.
% \end{align}
Clearly $\sum_i P_i = 1$. It is convenient to define $\vec{r}=\sum_{i}\vec{r}_i P_i$.
Then the energy polarization operator $\hat{\vec{R}}_E$ can be rewritten as  
\begin{align}
	\vec{R}_E &= \frac{1}{2}\sum_{\vec{R},\vec{R}'} \cre{c}_{\vec{R}}\qty((\vec{R}+\vec{r})H_{\vec{R}-\vec{R}'} + H_{\vec{R}-\vec{R}'} (\vec{R}'+\vec{r})) \ani{c}_{\vec{R}'} .
\end{align}
Then the total energy current operator is given by the time derivative of the energy polarization operator: $\hat{\vec{J}}^E = \dv*{\hat{\vec{R}}_E}{t}=(-i/\hbar)\comm{\hat{\vec{R}}_E}{\hat{H}}$. %, and the result is 
% \begin{align}
% 	\hat{\vec{J}}^E = \dv{\hat{\vec{R}}_E}{t} = -\frac{i}{\hbar}\comm{\hat{\vec{R}}_E}{\hat{H}}.
% \end{align}
The commutator can be written with $\ani{c}_{\vec{k}}, \cre{c}_{\vec{k}}, \hat{H}_{\vec{k}}$ as 
\begin{align}
	&\comm{\hat{\vec{R}}_E}{\hat{H}} %\nonumber\\ 
	=  \frac{1}{2}\sum_{\vec{R},\vec{R}',\vec{R}''} \cre{c}_{\vec{R}} \qty[(\vec{R}+\vec{r})H_{\vec{R}-\vec{R}'}H_{\vec{R}'-\vec{R}''} - H_{\vec{R}-\vec{R}'}H_{\vec{R}'-\vec{R}''}(\vec{R}''+\vec{r})] \ani{c}_{\vec{R}''} 
	% &=  \frac{1}{2N}\sum_{\vec{R},\vec{R}''} \cre{c}_{\vec{R}} \left[(\vec{R}+\vec{r}) e^{i\vec{k}\vdot(\vec{R}+\vec{r})}(H_{\vec{k}})^2e^{-i\vec{k}\vdot(\vec{R}''+\vec{r})} - e^{i\vec{k}\vdot(\vec{R}+\vec{r})}(H_{\vec{k}})^2e^{-i\vec{k}\vdot(\vec{R}''+\vec{r})} (\vec{R}''+\vec{r}) \right] \ani{c}_{\vec{R}''} \\
	% &= \sum_{\vec{k}} \cre{c}_{\vec{k}} i\nablak\qty[H_{\vec{k}}^2/2]\ani{c}_{\vec{k}}, \nonumber \\
	= \sum_{\vec{k}} \cre{c}_{\vec{k}} i\nablak\qty[H_{\vec{k}}^2/2]\ani{c}_{\vec{k}},
\end{align}
where $N$ is the total number of unit cells.
Therefore, the energy current operator $\hat{\vec{J}}^E$ is given by 
\begin{align}
	\hat{\vec{J}}^E &= \sum_{\vec{k}} \cre{c}_{\vec{k}} \frac{1}{2\hbar}\nablak\qty[H_{\vec{k}}^2]\ani{c}_{\vec{k}} 
	% \nonumber \\
	% =\hat{\vec{J}}^E &= \sum_{\vec{k}} \cre{c}_{\vec{k}} \frac{1}{2\hbar}\qty()\nablak\qty[H_{\vec{k}}^2]\ani{c}_{\vec{k}}
	. \label{eq:J_E}
\end{align}
Let us rewrite this expression with energy eigenstate basis. To this end, we introduce a unitary matrix $U_{\vec{k}}$ which diagonalizes the Bloch Hamiltonian $H_{\vec{k}}$.
\begin{align}
	E_{\vec{k}} &= U_{\vec{k}}^\dagger H_{\vec{k}} U_{\vec{k}}.
\end{align}
Here $E_{\vec{k}}$ is a diagonal matrix with its elements $(E_{\vec{k}})_{ll'} = \delta_{ll'}\ene_{\vec{k},l}$ where $\ene_{\vec{k},l}$ is the dispersion of the $l$-th band. Then the Hamiltonian $\hat{H}$ can be rewritten as 
\begin{align}
	\hat{H} &= \sum_{\vec{k}} \cre{d}_{\vec{k}} E_{\vec{k}} \ani{d}_{\vec{k}} = \sum_{\vec{k}, l} \ene_l(\vec{k})\cre{d}_{\vec{k},l}\ani{d}_{\vec{k},l},\\
	\ani{d}_{\vec{k}} &= U_{\vec{k}}^\dagger \ani{c}_{\vec{k}}.
\end{align}
Then $\hat{\vec{J}}^E$ is expressed with $\ani{d}_{\vec{k}},\cre{d}_{\vec{k}}$ as 
\begin{align}
	\hat{\vec{J}}^E &= \sum_{\vec{k}} \cre{d}_{\vec{k}} \frac{1}{\hbar}\qty(\nablak \qty[E_{\vec{k}}^2/2] - i\commutator{\vec{A}_{\vec{k}}}{E_{\vec{k}}^2/2}) \ani{d}_{\vec{k}}, \label{eq:J_E_operator}
\end{align}
where $\vec{A}_{\vec{k}}$ is the interband Berry connection defined as $\vec{A}_{\vec{k}}=U_{\vec{k}}^\dagger i\nablak U_{\vec{k}}$. 

As is well known, the electric current operator $\hat{\vec{J}}$ is given by
\begin{align}
	\hat{\vec{J}} &= \sum_{\vec{k}}\cre{c}_{\vec{k}}\frac{e}{\hbar}\nablak H_{\vec{k}} \ani{c}_{\vec{k}} = \sum_{\vec{k}} \cre{d}_{\vec{k}}\frac{e}{\hbar} \qty(\nablak E_{\vec{k}} - i\commutator{\vec{A}_{\vec{k}}}{E_{\vec{k}}}) \ani{d}_{\vec{k}}, \label{eq:J_operator}
\end{align}
with the charge of an electron $e$. Using $\hat{\vec{J}}^E$ and $\hat{\vec{J}}$, the heat current operator $\hat{J}^Q$ is defined as 
\begin{align}
	\hat{\vec{J}}^Q &= \hat{\vec{J}}^E - \frac{\mu}{e}\hat{\vec{J}}, \label{eq:J_Q_operator}
\end{align}
where $\mu$ is the chemical potential. The heat current operator can be written as 
\begin{align}
	\hat{\vec{J}}^Q &= \sum_{\vec{k}} \cre{c}_{\vec{k}} \frac{1}{2\hbar}\nablak\qty[(H_{\vec{k}}-\mu)^2]\ani{c}_{\vec{k}} = \sum_{\vec{k}} \cre{d}_{\vec{k}} \frac{1}{2\hbar}\qty(\nablak \qty[(E_{\vec{k}}-\mu)^2] - i\commutator{\vec{A}_{\vec{k}}}{(E_{\vec{k}}-\mu)^2}) \ani{d}_{\vec{k}}. \label{eq:J_Q}
\end{align}

\section{Calculation of correlation functions} \label{Ap:calc_linres}
Here we evaluate the correlation functions Eq.~\eqref{eq:corr_func_green}.
The summation over the Matsubara frequency $\epsilon_n$ can be taken by the usual technique. 
The first term in Eq.~\eqref{eq:corr_func_green}, the intraband contribution, reads
\begin{align}
	&-\kB T\sum_n \G_{l}(i\ene_n, \vec{k})\G_{l}(i\ene_n-i\omega_\lambda, \vec{k})
	\nonumber \\
	&\xrightarrow{i\omega_\lambda\to \omegap} \int\frac{\dd{\epsilon}}{2\pi i}f(\epsilon)(\GR_l(\epsilon,\vec{k})-\GA_l(\epsilon, \vec{k})) (\GR_l(\epsilon+\omegap,\vec{k}) + \GA_l(\epsilon-\omegap,\vec{k})) \\
	&= \int\frac{\dd{\epsilon}}{2\pi i}f(\epsilon)(\GR_l(\epsilon,\vec{k})^2-\GA_l(\epsilon, \vec{k})^2)
	+ \omegap\int\frac{\dd{\epsilon}}{2\pi i}f(\epsilon)\pdv{\epsilon}\qty[\GR_l(\epsilon,\vec{k})-\GA_l(\epsilon, \vec{k})] (\GR_l(\epsilon,\vec{k})-\GA_l(\epsilon, \vec{k})) \nonumber \\ 
	&+ \frac{{\omegap}^2}{2} \int\frac{\dd{\epsilon}}{2\pi i}f(\epsilon)\pdv[2]{\epsilon}\qty[\GR_l(\epsilon,\vec{k})+\GA_l(\epsilon, \vec{k})] (\GR_l(\epsilon,\vec{k})-\GA_l(\epsilon, \vec{k})) + \order{\omegap^3} \\
	% &-f'(\xi_{\vec{k},l})(1 + i\omega\tau_{\vec{k},l} - (\omega\tau_{\vec{k},l})^2 + \order{\omegap^3}) \\
	&= \int\frac{\dd{\epsilon}}{2\pi i}f(\epsilon) \qty(2i\Im{(\GR_l)^2}  -2\omegap\pdv{\epsilon}\qty[(\Im\GR_l)^2] + 2i\omegap^2\Im\GR_l\pdv[2]{\epsilon}\qty[\Re\GR_l]) + \order{\omegap^3} 
\end{align}
where $f(\epsilon) = (e^{\beta\epsilon}+1)^{-1}$ is the Fermi distribution function, $f'(\epsilon) = \dd{f}/\dd{\epsilon}$, $\GR_l(\epsilon, \vec{k}) = (\epsilon-\xi_{\vec{k},l} + i\Gamma_{\vec{k},l})^{-1}$ is the retarded Green's function, and $\omegap=\omega+i\eta$. In the last line, the arguments of $\GR_l$ are omitted.  
The integrand can be rewritten as follows.
\begin{align}
	&2i\Im{(\GR_l)^2}  -2\omegap\pdv{\epsilon}\qty[(\Im\GR_l)^2] + 2i\omegap^2\Im\GR_l\pdv[2]{\epsilon}\qty[\Re\GR_l] \nonumber \\
	&= \pdv{\epsilon}\left[\frac{2i\Gamma_{\vec{k},l}}{(\epsilon-\xi_{\vec{k,l}})^2 + \Gamma_{\vec{k},l}^2} - \omegap \frac{2\Gamma_{\vec{k},l}^2}{((\epsilon-\xi_{\vec{k},l})^2 + \Gamma_{\vec{k},l}^2)^2} + 2i\omegap^2 \qty(-\frac{4\Gamma_{\vec{k},l}^3}{3}\frac{1}{((\epsilon-\xi_{\vec{k},l})^2 + \Gamma_{\vec{k},l}^2)^3} + \frac{\Gamma_{\vec{k},l}}{2}\frac{1}{((\epsilon-\xi_{\vec{k},l})^2 + \Gamma_{\vec{k},l}^2)^2}) \right]
\end{align}
When $\Gamma_{\vec{k},l}$ is small, the functions in the square bracket are appreciable only when $\epsilon\simeq\xi_{\vec{k},l}$ and thus can be approximated with the delta-function.
\begin{align}
	&\frac{1}{(\epsilon-\xi_{\vec{k,l}})^2 + \Gamma_{\vec{k},l}^2} \simeq \frac{\pi}{\Gamma_{\vec{k},l}} \delta(\epsilon-\xi_{\vec{k},l}) \\
	&\frac{1}{((\epsilon-\xi_{\vec{k},l})^2 + \Gamma_{\vec{k},l}^2)^2} \simeq \frac{\pi}{2\Gamma_{\vec{k},l}^3} \delta(\epsilon-\xi_{\vec{k},l})\\
	&\frac{1}{((\epsilon-\xi_{\vec{k},l})^2 + \Gamma_{\vec{k},l}^2)^3} \simeq \frac{3\pi}{8\Gamma_{\vec{k},l}^5} \delta(\epsilon-\xi_{\vec{k},l})
\end{align}
The numerical factors in the right-hand side are determined so that the integration over $[-\infty, \infty]$ coincides with that of the left-hand side. 
Then the Matsubara summation after the analytic continuation reads
\begin{align}
	&-\kB T\sum_n \G_{l}(i\ene_n, \vec{k})\G_{l}(i\ene_n-i\omega_\lambda, \vec{k})
	\nonumber \\
	&\xrightarrow{i\omega_\lambda\to \omegap} \int\dd{\epsilon}f(\epsilon) \qty(1+i\omegap\tau_{\vec{k},l} - \omegap^2\tau_{\vec{k},l}^2 + \order{\omegap^3}) \pdv{\ene}\delta(\ene-\xi_{\vec{k},l}) \\
	&= -f'(\xi_{\vec{k},l}) \qty(1+i\omegap\tau_{\vec{k},l} - \omegap^2\tau_{\vec{k},l}^2 + \order{\omegap^3}) 
\end{align}
The second term in Eq.~\eqref{eq:corr_func_green}, which represents the interband contribution, reads
\begin{align}
	&-\kB T\sum_n \G_{l'}(i\epsilon_n, \vec{k})\G_{l'}(i\epsilon_n-i\omega_\lambda, \vec{k}) \nonumber \\
	&= \int_{-\infty}^{\infty}\frac{\dd{\epsilon}}{2\pi i}f(\epsilon)\left[G_{l'}^R(\epsilon+i\omega_\lambda)(G_l^R(\epsilon)-G_l^A(\epsilon)) + (G_{l'}^R(\epsilon)-G_{l'}^A(\epsilon))G_l^A(\epsilon-i\omega_\lambda) \right] \\
	&\simeq \int_{-\infty}^{\infty}\dd{\epsilon}(-f(\epsilon))\left[ G_{l'}^R(\epsilon+i\omega_\lambda)\delta(\epsilon-\xi_{\vec{k},l}) + \delta(\epsilon-\xi_{\vec{k},l'})G_l^A(\epsilon-i\omega_\lambda) \right] \\
	&= -f(\xi_{\vec{k},l})G_{l'}^R(\xi_{\vec{k},l}+i\omega_\lambda) -f(\xi_{\vec{k},l'})G_l^A(\xi_{\vec{k},l'}-i\omega_\lambda) \\ 
	&= -\frac{f(\xi_{\vec{k},l})}{\xi_{\vec{k},l} - \xi_{\vec{k},l'} + i\omega_\lambda + i\Gamma_{\vec{k},l'}}+\frac{f(\xi_{\vec{k},l'})}{\xi_{\vec{k},l} - \xi_{\vec{k},l'} + i\omega_\lambda + i\Gamma_{\vec{k},l}} \\
	& \xrightarrow{i\omega_\lambda\to \omegap} -\frac{f(\xi_{\vec{k},l})}{\xi_{\vec{k},l} - \xi_{\vec{k},l'} + \omegap + i\Gamma_{\vec{k},l'}}+\frac{f(\xi_{\vec{k},l'})}{\xi_{\vec{k},l} - \xi_{\vec{k},l'} + \omegap + i\Gamma_{\vec{k},l}}
\end{align}
where the argument $\vec{k}$ of $\GR_l$ is omitted.
We have used an approximation, $G_l^R(\epsilon)-G_l^A(\epsilon)=2i\Gamma_{\vec{k},l}/((\epsilon-\xi_{\vec{k},l})^2 + \Gamma_{\vec{k},l}^2)\simeq 2\pi i \delta(\epsilon-\xi_{\vec{k},l})$. 
If the band is gapped and the band gap is large enough so that $\xi_{\vec{k},l}-\xi_{\vec{k},l'} \gg \Gamma_{\vec{k},l}$, $\Gamma_{\vec{k},l}$ in the denominator can be neglected and we obtain
\begin{align}
	&-\kB T\sum_n \G_{l'}(i\ene_n, \vec{k})\G_{l'}(i\ene_n-i\omega_\lambda, \vec{k}) \xrightarrow{i\omega_\lambda\to \omegap} \frac{f(\xi_{\vec{k},l'})-f(\xi_{\vec{k},l})}{\xi_{\vec{k},l} - \xi_{\vec{k},l'} + \omegap}
\end{align}
Therefore, the correlation functions become
\begin{align}
	\Phi_{ij}(\omega) &= -\frac{1}{V}\sum_{\vec{k}}\sum_{l}(J_{i, \vec{k}})_{l,l}(J_{j, \vec{k}})_{l,l} f'(\xi_{\vec{k},l}) \qty(1+i\omegap\tau_{\vec{k},l} - \omegap^2\tau_{\vec{k},l}^2 + \order{\omegap^3}) +\frac{1}{V}\sum_{\vec{k}, n} \sum_{l\neq l'}(J_{i, \vec{k}})_{l,l'}(J_{j, \vec{k}})_{l',l} \frac{f(\xi_{\vec{k},l'})-f(\xi_{\vec{k},l})}{\xi_{\vec{k},l} - \xi_{\vec{k},l'} + \omegap}
\end{align}
By expanding the second term with respect to $\omega$ and substituting into Eq.~\eqref{eq:response_function}, we obtain
\begin{align}
	&L_{ij}(\omega) = -\frac{1}{V}\sum_{\vec{k},l}(J_{i, \vec{k}})_{l,l}(J_{j, \vec{k}})_{l,l} f'(\xi_{\vec{k},l})\tau_{\vec{k},l}(1 +i \omegap\tau_{\vec{k},l} + \order{\omegap^2}) 
	-\frac{1}{V}\sum_{\vec{k}, l\neq l'}(J_{i, \vec{k}})_{l,l'}(J_{j, \vec{k}})_{l',l} \frac{f(\xi_{\vec{k},l})-f(\xi_{\vec{k},l'})}{\qty(\xi_{\vec{k},l} - \xi_{\vec{k},l'})^2} i\qty(1-\frac{\omegap}{\xi_{\vec{k},l}-\xi_{\vec{k},l'}} + \order{\omegap^2}) \label{eq:Lij_general}
\end{align}
By substituting Eqs.~\eqref{eq:J_E_operator}, \eqref{eq:J_Q_operator}, and \eqref{eq:J_operator} into Eq.~\eqref{eq:Lij_general}, we obtain the results Eqs.~\eqref{eq:L11_Kubo} and \eqref{eq:L12_Kubo}. In Eqs.~\eqref{eq:L11_Kubo} and \eqref{eq:L12_Kubo}, $\omegap=\omega+i\eta$ is replaced by $\omega$ since $\eta$ is usually an infinitesimal quantity. 

\section{Calculation of SSH model}
Here we present the detail of the calculation of the SSH model. The SSH model is given by
\begin{align}
	H = \sum_n (t_{1} \cre{c}_{nB} \ani{c}_{nA} + t_{2} \cre{c}_{n+1,A} \ani{c}_{nB} + \mathrm{h.c.}) + \frac{\Delta}{2} (\cre{c}_{nA} \ani{c}_{nA} - \cre{c}_{nB} \ani{c}_{nB}) \label{ap:eq:SSH}
\end{align}
where $\ani{c}_{n{\alpha}}$ is an annihilation operator for site $\alpha=A,B$ in $n$-th unit cell, each localted at $na+r_{\alpha}$ with lattice constant $a$. Site $A$ and $B$ in the same unit cell are connected by the hopping $t_{1}$, and the hopping between sites $A$ and $B$ in neighboring unit cells is $t_{2}$. The onsite energy difference between the sites $A$ and $B$ is $\Delta$. 

We define the Fourier transformation of the annihilation operator as
\begin{align}
	\ani{c}_{k\alpha} &= \frac{1}{\sqrt{N}}\sum_n e^{-ik(na+r_{\alpha})} \ani{c}_{n\alpha},
\end{align}
where $N$ is the total number of unit cells. Then the Bloch Hamiltonian is given by
\begin{align}
	H = \begin{pmatrix}
		\cre{c}_{kA} & \cre{c}_{kB}
	\end{pmatrix}
	\begin{pmatrix}
		\Delta/2 & (t_1 + t_2 e^{-ika})e^{-ik(r_A-r_B)} \\
		(t_1 + t_2 e^{ika})e^{ik(r_A-r_B)} & -\Delta/2
	\end{pmatrix}
	\begin{pmatrix}
		\ani{c}_{kA} \\
		\ani{c}_{kB}
	\end{pmatrix}
\end{align}
Then the interband Berry connection $\vec{A}_{nm}$ is given by 
\begin{align}
	\vec{A}_{nm} &= i\mel{u_n}{\nablak}{u_m} = -i\frac{\mel{n}{\nablak H}{m}}{\ene_n-\ene_m}
\end{align} 
with $\ket{u_n}$ the cell-periodic Bloch wavefunction for $n$-th band. With this expression of the Berry connection, we numerically calculate the thermopolarizability given by 
\begin{align}
	\chit^{\alpha\beta} = -e \sum_n \int[\dd{k}] G_n^{\alpha\beta} & \qty((\ene_n-\mu)f(\ene_n-\mu)  + \kB T \log[1+e^{-\beta(\ene_n-\mu)}]),  \label{ap:eq:thermopolarizability1} \\
	G_n^{\alpha\beta} &= -2\sum_{m(\neq n)} \frac{\Re[A^\alpha_{nm}A^\beta_{mn}]}{\ene_n-\ene_m}. \label{ap:eq:Gn}
\end{align}
The results are shown in Fig. 3 in the main text.
\end{widetext}

\end{document}